\documentclass[aps,12pt]{revtex4}

\newcommand{\be}{\begin{eqnarray}}
\newcommand{\ee}{\end{eqnarray}}
\usepackage{graphicx,epsf}    
\begin{document}
 
\columnsep     38pt
\oddsidemargin  5pt
\parsep  3pt plus 1pt minus 1pt
\title{\bf Final state interaction in lepton scattering off deuterons}

\author{{L.P. Kaptari}}       
\altaffiliation[Also at ]{ INFN, Sezione di Perugia,  via A. Pascoli, Perugia, I-06100, Italy}         
\affiliation{ Bogoliubov Lab. Theor. Phys., JINR, 141980  Dubna, Russia }

\date{\today} 

\begin{abstract}
\vspace*{0.5cm}   
The role of the  final state interactions (FSI) in inclusive electro disintegration
of the deuteron,  
$D(e,e')X$,  is being investigated within different approaches.
  A detailed comparison between 
 an improved Glauber method  and the standard Schroedinger 
 approach is presented.
It is shown that both methods become inadequate at large  values of $Q^2$, 
where the virtuality of the hit nucleon after 
photon absorbtion is very high.  
The concept of finite
formation time (FFT) required by the hit hadron to reach its asymptotic form
 is introduced  by  a  Feynman diagram approach and  by explicitly 
 taking into account  the dependence of the 
  ejected nucleon on its virtuality.
 The approach has been  applied both at  $x_{Bj}\simeq 1$,
as well as in the  so called {\it cumulative} region i.e. at
$x_{Bj} > 1$. Numerical calculations  
 show that the effects of the  FFT  almost completely  
cancel the contribution from rescattering processes.
In the cumulative region  the color transparency or finite formation
time effects become fairly visible.
\end{abstract}
\maketitle

\section{Introduction}

  One of the most challenging  aspects of the nowadays   hadronic 
  physics is the experimental and theoretical investigation
  of effects of Final State Interaction (FSI) in electromagnetic processes off
  nuclei at high energy and momentum transfers. 
  As a matter of fact, these processes are viewed as a useful tool 
  for studying  several features 
 of  strong interactions at short distances  and 
for  checking  the   predictions of   perturbative QCD.
At high moment  transfer, when the electron probes extremely short distances, 
 the strong interaction is  expected to be screened in the same way as the atomic 
 Coulomb field is screened at short distances  in QED. 
  In other words, in the quark-gluon debris, produced    by the target quark which 
 absorbed the highly virtual  photon, the color forces are screened
 ("switched off")  in the neighbourhood of the interaction spot. 
 Only at a certain  distance from the interaction point, 
 known as the  Formation or Hadronization Length
 (or, equivalently, the Finite Formation Time),
 the reaction products become able to interact. Moreover,  
 the magnitude of the Hadronization Length strongly depends upon the kinematics of
 the process and, at least theoretically, there can exist situations when
 the strong interaction may completely vanish, a phenomenon known as 
 Color Transparency (see for detail Refs. \cite{zam,ct} and references
 therein quoted).  From the point of view of  nuclear physics  
this means that after absorbtion of $\gamma^*$ the quark system is 
able to produce many physical states which,  due to their  coherent interaction  in
the final state, manifest a strong attenuation of the cross section with respect to the
prediction of the usual  eikonal theory. In the presence of CT, the ratio of the  
experimentally  observed cross section to the one calculated theoretically
within the Plane Wave Impulse Approximation (PWIA) is predicted  to approach unity as
$Q^2\to \infty$.   However, at finite $Q^2$ one should investigate the ratio of
experimentally measured quantities to the ones computed  taking into account the
FSI effects;  an observed deviation of such a  ratio from unity,  
in agreement with the $Q^2$ dependence  predicted by CT,
would be  unambiguous signal of the latter.  
Presently, evidence  from the experimental study of proton
production in $eA$ collisions seems to indicate that the CT effects
are not yet visible up to $Q^2$ of the order $20\  GeV^2/c^2$, which means
that in this region  the experimental data can satisfactorily be described within usual
theoretical approaches. Since CT corrections
are predicted to be small at finite $Q^2$, and  because of the required  theoretical 
difficulties in the treatment of FSI and CT effects, the choice of the
processes and kinematics to be investigated,  as well as of  the theoretical frameworks for the  interpretation
of the data, is  of  great importance.  

 Another noteworthy aspect   of FSI effects is its tight connection with 
 the problem of the experimental investigation   
 of   Ground State Correlations (GSC)  in nuclei, particulary those 
 which originate from the $NN$-correlations 
 at short distances. There is  a renewed interest in this topic
 in view of  new experimental  possibilities
   in the investigation of electro disintegration of  light nuclei  
 in semi inclusive and exclusive reactions  (about the present status of 
 experimental and theoretical investigations 
 of GSC in nuclei see, e.g., Ref \cite{lund}). 
 A detailed study of GSC  would  allow to obtain additional  
 information about the nuclear wave function and
 to directly check the validity of the Standard Model of nuclei as systems of
 non relativistic nucleons interacting via known $NN$ forces.  
 Unfortunately, contrary to elastic processes,  these inelastic electromagnetic reactions 
 are affected by several effects ( Final State Interaction
 among the reaction products, Meson Exchange Currents effects etc.),
 which may mask the effects of GSC. Therefore one needs reliable  methods
 of calculations of the FSI corrections in these processes as well. 
 Besides, since the study of  GSC at short distances  requires  probes with relatively high 
 momentum transfer, the question of the role of relativistic corrections cannot be neglected 
 in theoretical treatments of the mechanism of the electro disintegration of nuclei.  
 
 In this context, the problem of a reliable theoretical calculations of
  relativistic effects, FSI  and FFT effects appears to be of a great  interest. 
 In  the present talk a study of the  mentioned effects in the inclusive process
 of electro disintegration of the deuteron $D(e,e')X$,
 is presented.  Investigations of the semi inclusive
 reactions of the type $A(e,e'p)$ for deuterons and light nuclei is in progress and will
 be reported elsewhere \cite{tolpa}.

\section{Kinematics and cross section}

 \begin{figure}[th]
\raggedright
 \includegraphics[height=6.cm]{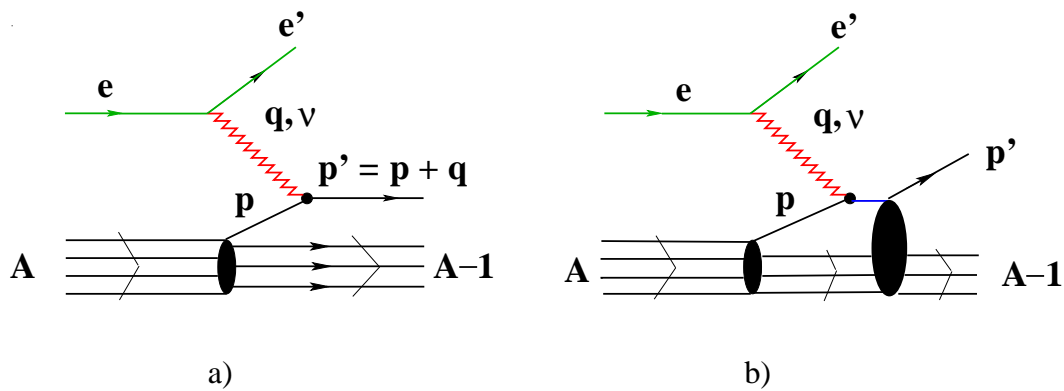}
\caption{ The diagrams of quasi elastic
electro disintegration of a nucleus $A$ within  the PWIA (a)
and  taking into account the  Final State Interaction (b).
}
\label{fig1}
\end{figure}
The invariant cross section for the electro disintegration of the target $A$ into the
final hadron system  $H$,
  can be written in a general form as

\begin{eqnarray}
d\sigma = \frac{1}{4kp_A}\left|  {\cal M}_{e+A \to e' + H }\right |^2
(2\pi)^4\delta^{(4)}\left( k+p_A -k'- p_{H} \right) \frac{d{\bf  k'}}{(2\pi)^3 2{\cal E}'}
d\tau_H,
\label{cs00}
\end{eqnarray}
 where
the
square of the invariant matrix element
$\left|  {\cal M}_{e+A \to e' + H }\right |^2 $
in the one-photon-exchange approximation can be written as  
\begin{eqnarray}
 \left|  {\cal M}_{e+A \to e' + H }\right |^2 (2\pi)\delta^{(4)}((p_1+q)^2-m^2)
= \frac{ e^4}{Q^4} L^{\mu\nu}(k,q) W_{\mu\nu}^A(p_A,q),
\label{ime00}
\end{eqnarray}
with the known electromagnetic leptonic tensor $L^{\mu\nu}(k,q)$,
  $L^{\mu\nu}(k,q) = 2 \left(
  2k_\mu k_\nu -(k_\mu q_\nu +k_\nu q_\mu) + g_{\mu\nu}\frac{q^2}{2}
  \right)$.
  The (generally unknown) 
  hadronic tensor $W^A_{\mu\nu}$  besides electromagnetic
  interaction also involves information about the strong structure of the interaction currents.
  So, in case of elastic electron scattering from nucleons the
  electromagnetic nucleon vertex $\Gamma_\mu^{eN}(Q^2)$ is 
   usually  parametrized  in terms of two elastic form factors
 \begin{eqnarray}
\Gamma^{eN}_{\mu}(Q^2) = \gamma_{\mu} F_1(Q^2) +
i \frac{\sigma_{\mu\alpha}q^{\alpha}}{2m}\kappa F_2(Q^2),
\label{vertex}
\end{eqnarray}
with the resulting tensor $W^N_{\mu\nu}$  
\begin{eqnarray}
 W^N_{\mu\nu}(p_1,q)&& = \frac{1}{2} {\sf Tr} \left \{
 (\hat p_1 +m)\Gamma_\mu^{eN}(Q^2)  (\hat p_1+\hat q +m)
\Gamma_\nu^{eN}(Q^2)
 \right \}(2\pi)\delta\!\left((p_1+q)^2-m^2\right),
\label{ime02}
 \end{eqnarray}
 which being contracted  with the leptonic tensor provides
 the well known Rosenbluth formula for the cross section.
 
In the case of a complex system the form of
the hadronic tensor cannot be unambiguously defined and some theoretical models
have to be used to relate the nuclear tensor via the known nucleon one.
In what follows, for the sake of simplicity  we concentrate attention
on the simplest nucleus, the deuteron. However, all results and conclusions
 hold for complex nuclei as well.

\subsection{PWIA}
The simplest theoretical approach to calculate the hadronic tensor is the
plane wave impulse approximation (PWIA) which   assumes that
i)  the nuclear electromagnetic current is a sum of solely  one-nucleon currents 
ii) the final state interaction between the struck nucleon and the residual 
nucleus does not contribute to the cross section (see fig. \ref{fig1}a). 
The neglect of FSI
implies that the state vector describing the final system can be presented
as a tensor product of the corresponding vectors of the struck nucleon and the residual nucleus.
 This immediately allows to separately
 compute  the matrix elements of nucleon currents and the 
 part describing the   nuclear  $A-1$ structure. 
 A direct calculation of the diagram on fig.\ref{fig1}a
 (in both non relativistic \cite{ck} and relativistic \cite{farrali}) approaches
 leads to  the  following form of the inclusive cross section    
\begin{equation}
\left ( \frac{d\sigma}{d{{\cal E}'}d\Omega_{k'}} \right )^{PWIA}_{eD}
=
 (2\pi)\,   \int\limits_{p_{min}}^{p_{max}}    |{\bf  p}|  \, d|{\bf  p}|
 n^D({\bf  p})  \frac{E_{\vec p+\vec q} }{|{\bf q}|}
 \left (\overline\sigma_{ep}+\overline\sigma_{en}\right ),
 \label{pwia}
 \end{equation}
 where $ \left (\overline\sigma_{ep}+\overline\sigma_{en}\right )$ represents
 the cross section of electron scattering from an isoscalar 
 nucleon; the quantity $ n_D$ in the non relativistic limit
 is  the familiar deuteron momentum distribution related with the deuteron wave function
 as follows 
 \begin{equation}
 n^{D}(|{\bf p}|) =\frac{1}{3(2\pi)^3} \sum\limits_{{\cal M}_D}
\left |\int \Psi_{{1,\cal M}_D} ({\bf r})\ \exp (i{\bf
p}{\bf r}) d{\bf r} \right |^2. \label{deutwf}
\end{equation} 
\begin{figure}[ht]
\raggedleft
\begin{minipage}{3in}
 \includegraphics[height=0.3\textheight]{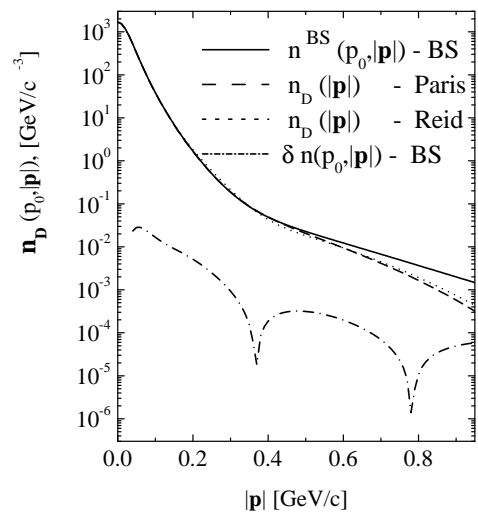}
 \end{minipage}
 
 \raggedright
 \vskip -2.5in
 \begin{minipage}{3in}
\caption{{\it The deuteron momentum distribution computed within the covariant
Bethe-Salpeter formalism (solid line) in comparison with
non relativistic calculations with realistic $NN$-interactions, the Paris-group
(dashed line) and Reid Soft Core (dotted line ) potentials. The relativistic corrections
from the $P$-waves in the deuteron are represented by 
the dash-dotted line (cf. Ref.\cite{farrali}).}}
\label{fig2}
\end{minipage}
\vskip 1cm
\end{figure}

In the relativistic case an analogue 
of the momentum distribution can be introduced
as well, and in terms of the Bethe-Salpeter partial amplitudes reads as \cite{farrali}
\begin{equation}
{n}^{BS}_D(p_0,{\bf p})= 
\left (
\Psi_S^2(p_0,|{\bf p}|)+\Psi_D^2(p_0,|{\bf p}|) \right )+\frac{|{\bf p}|}{E_{{\bf p}}}\delta {\rm N}(p_0,{\bf p}),
\label{distrBS}
\end{equation}
where the contribution from pure relativistic corrections ("negative" P-waves)  is 
\begin{eqnarray}
 &&
\delta { n}(p_0,{\bf p})=
\left \{
\frac{2\sqrt{3} }{3}
\left [ \phantom{\frac 11}
\Psi_S(p_0,|{\bf p}|) \left(\Psi_{P_1}(p_0,|{\bf p}|) - \sqrt{2} \Psi_{P_3}(p_0,|{\bf p}|)
\right ) + \right. \right.
\nonumber\\[1mm]
&&
\hspace*{3cm}\left.\left.\phantom{\frac 12}
\Psi_D(p_0,|{\bf p}|) \left( \sqrt{2}\Psi_{P_1}(p_0,|{\bf p}|) +\Psi_{P_3}(p_0,|{\bf p}|)
\right) \right ] \right \}.
\label{deltan}
\end{eqnarray}
Now the problem of whether the relativistic corrections play a role in the considered
processes reduces to a comparison of the  momentum   distributions, 
computed within relativistic and
the non relativistic approaches. 
In Fig. \ref{fig2} the 
non relativistic deuteron momentum distribution predicted by realistic
 potentials (RSC and Paris-group)
are compared with the ones  obtained within 
the covariant BS formalism with a  realistic one-boson-exchange
kernel \cite{ku}. It can be  seen that the relativistic 
corrections (\ref{deltan}) are negligibly small
so that in the kinematical region of our interests  the two approaches
provide basically identical results. 
Thus one can conclude that relativistic correction are irrelevant in the 
considered kinematical region and, for the  sake of simplicity, 
all further analysis can be safely performed within
the non relativistic limit.   
It is also worth emphasizing that corrections like relativistic  meson exchange
currents can be neglected as well, for in the covariant BS calculations of the momentum distribution
the $N\bar N$- pair 
production is already taken into account  within the PWIA \cite{semikh}.

\section{The rescattering contribution}

\subsection {The  Schroedinger Approach}
At moderate  momentum transfer $Q^2$ the total energy of the ejectile and the $A-1$
system can be  relatively small, below the pion production threshold. 
In this case the most reliable
method to take into account the FSI effects is to 
use the solution of  the Schroedinger equation in 
the continuum for the final system.
The corresponding cross section for the
longitudinal current is \cite{ck}   
\begin{eqnarray}
{d^3 \sigma^L \over  d \Omega' d {\cal E}'
}=\frac{4}{3} \frac{ M^2\, \sigma_{Mott} }{2\pi}
V_L\, G_E(Q^2)^2
\sum\limits_{J_f}\sum\limits_\lambda \left |
\langle J_D|| \hat O_\lambda(|{\bf q}|)
||p_{rel}; J_f L_f S_f \rangle \right |^2
 \frac{|{\bf p}_{rel}|}{\sqrt{s}}.
\label{wigner}
\end{eqnarray}
where  ${\bf p}_{rel}$ is the relative momentum of
the $np$-pair which  is defined by the  Mandelstam variable
 $s=4\,({\bf p}_{rel}^2+M^2)$;
 ${\rm \hat O_{\lambda \mu}}=  i^\lambda j_\lambda
(qr/2) Y_{\lambda
\mu}(\hat r)$
 and   the scattering state  $| p_{rel};J_f L_f S_f \rangle  $ 
 describes the interaction of the two nucleons  in the continuum.

It can be seen that  equation (\ref{wigner})
differs  from the PWIA result (\ref{pwia}).
 However, by using the
identity $\displaystyle\frac{1}{2|{\bf
q}|}\displaystyle\int\limits_{|y|}^{p_{max}}
\displaystyle\frac{| {\bf p}| d | {\bf
p}|}{E}=\frac{p_{rel}}{\sqrt{s}}$ one may  cast
the cross section in the form of eq. (\ref{pwia}) by 
formally replacing the momentum distribution 
$n_D(|{\bf p}|)$ 
with  the quantity 
$ n_S^D(|{\bf  p}|, |\textbf{q}|, \nu)$,   
\begin{equation}
 n_S^D(|{\bf  p}|, |\textbf{q}|, \nu)
=\frac{1}{4\pi}\,\frac{1}{3}\sum\limits_{J_f}\sum\limits_\lambda
\left | \langle J_D|| \hat O_\lambda(|{\bf q}|)
|| p_{rel};J_f L_f S_f \rangle \right |^2,
\label{newdist}
\end{equation}
which can be called the distorted momentum distribution within the Schroedinger
approach. Note, that in absence of FSI, the state vector  
$| p_{rel};J_f L_f S_f \rangle  $ is nothing but the multipole 
decomposition of the $np$ plane wave and eq. (\ref{newdist}) 
coincides exactly with the usual   PWIA momentum distribution , eq. (\ref{deutwf}).   

\subsection {The Glauber Approach}
In the Glauber approach the exact two-nucleon continuum wave function
$|f>$ is approximated by its  eikonal  form.
Then the cross section can be written in the same
form as equation (\ref{pwia}) with the deuteron momentum
distribution (\ref{deutwf}) replaced by the Glauber distorted
momentum distribution $n_G^D$ \cite{ck},
\begin{equation}
n^D( {\bf p}) \to n_G^D( {\bf p}_m) =
\frac13\frac{1}{(2\pi)^3} \sum\limits_{{\cal
M}_D} \left | \int\, d  {\bf r} \Psi_{{1,\cal
M}_D}^*( {\bf r}) S( {\bf r}) \chi_f\,\exp (-i
{\bf p}_m {\bf r}) \right |^2, \label{ddistr}
\end{equation}
where
$ 
{\bf p}_m = {\bf q}-{\bf p}_1' 
$
is the missing momentum,
 $\chi_f$  the spin wave
function of the final $np$-pair  and $S( {\bf r})$
 the $S$-matrix describing the  final state interaction
between the hit nucleon and the spectator, 
\begin{equation}
S({\bf r}) = 1-\theta(z)\,\Gamma_{el}({\bf b}),
\label{sg}
\end{equation}
where  $\Gamma_{el}({\bf b})$ is the elastic profile function,
$
\Gamma_{el}({\bf b})=\displaystyle\frac{\sigma_{tot}(1-i\alpha)}{4\pi b_0^2}\,
\exp(-b^2/2b_0^2) $, and
  ${\bf r} =
{\bf b} + z\, {\bf q}/|{\bf q}|$ defines the
longitudinal, $z$, and the  perpendicular, ${\bf b}$,
components of the relative coordinate {\bf r},
$\sigma_{tot}=\sigma_{el} + \sigma_{in}$, $\alpha$ is
 the ratio of the real to the
imaginary part of the forward elastic $pn$ scattering
amplitude, and, eventually,  the step function $\theta(z)$
originates from the Glauber's high energy
approximation, according to  which the struck nucleon
propagates along a straight-line trajectory and
can interact with the spectator only provided $z
> 0$.

 \begin{figure}[t]
\raggedright
 \hspace*{-1cm}\includegraphics[height=6cm]{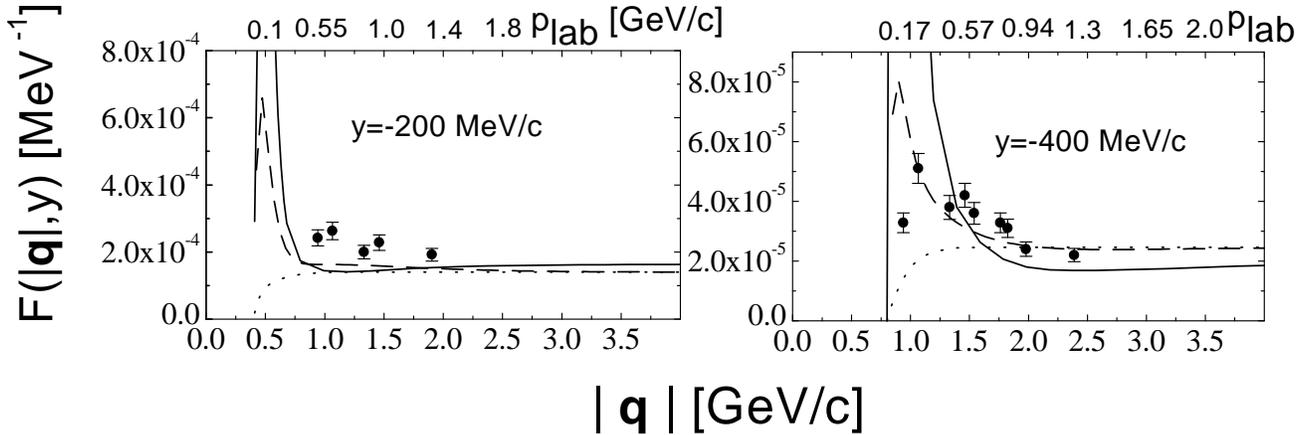}
\caption{\it The scaling function $F(|{\bf q}|,y)$ as a 
function of $|{\bf q}|$ at two values of
 $|y|=p_{min}$, eq. (\ref{pwia}).
Dotted lines - PWIA. 
Dashed lines - FSI taken into 
account by the  Schroedinger approach;
full lines - FSI taken into 
account by the Glauber approximation (c.f. Ref. \cite{ck})
}
\label{fig3}
\end{figure}
It should be stressed, first, that in absence of any FSI,
the distorted momentum distribution  $n_G^D({\bf
p}_m)$ reduces to the undistorted momentum distribution
 $n^D({\bf p})$  (${\bf p}_m =-{\bf p}_1$) and, secondly,  that unlike   $n^D({\bf
p})$,   $n_G^D({\bf
p}_m)$ depends also upon the orientation
of ${\bf  p}_m$ with respect to the momentum transfer
 ${\bf  q}$, with  the angle $\theta_{{\bf qp}_m}$ being fixed by the
 energy conserving $\delta$-function, namely
 $\cos \theta_{{\bf qp}_m}=[(2(M_D+\nu)\sqrt{|\textbf{p}_m|^2+M^2}
 -s)]/(2|\textbf{q}||\textbf{p}_m|)$;
 thus
 $n_G^D({\bf  p}_m)$
depends implicitly on the kinematics of the process, and
the values of  $p_{min}$ and  $|{\bf  q}|$
fix the value of the total energy
  of the final $np$ pair, i.e. the
relative energy of the nucleons in the final
states. Consequently, the quantities
$\sigma_{tot}$, $\alpha$ and $b_0$ in
  also depend upon the kinematics of
the process. In this sense, the distorted
momentum distribution $n_G^D({\bf  p}_m)$ implicitly
depends upon $|{\bf  q}|$ and $p_{min}$ as well.

\subsection{Schroedinger vs Glauber: numerical results}

A formal comparative analysis of FSI effects can be performed
by comparing the corresponding momentum distributions eqs.( \ref{newdist}),
(\ref{ddistr}) with the PWIA results, eq. (\ref{deutwf}). However such a direct 
comparison may not be quite enlightening, for in the considered processes besides
the momentum distribution, $n^D$, the cross section
is also determined by other kinematical factors and restrictions, 
which can strongly influence the magnitude of FSI effects. For this reason
it is more convenient to 
 present  the results of 
 calculation in terms of  the scaling function $F({\bf q},y)$, which is 
obtained from the cross section by eliminating all the irrelevant 
 kinematical factors.
 In Fig. \ref{fig3}
  the PWIA (dotted curves) is compared with the results which
 include  FSI effects within
 the Schroedinger approach (dashed curves) and the Glauber approximation (solid curves).
 The quantity $|y|=p_{min}$, eq. (\ref{pwia}), determines which part of the
 intrinsic nucleon momenta contributes to the cross section. At low total energies $\sqrt{s}$
 of the $np$ pair ($s=2m^2+2m\sqrt{m^2+p_{lab}^2}$) the Schroedinger approach
 provides a rather good description of the  data, while the Glauber approximation is  not
 valid at such relative energies. 
 With  increasing momentum   transfer, the FSI contribution,
   computed within both approaches, decreases. Note that above the pion
   production threshold the use of the Schroedinger approach becomes  questionable.
   Note also that, as seen from fig.\ref{fig3}, at high  momentum transfer the FSI corrections
   computed within the Glauber approximation  remain almost constant.

\section{Finite Formation Time}
As illustrated in  the previous section, there is only a limited interval of 
 momentum transfer $Q^2$ where the Glauber approximation holds: at  low
relative energy of the interacting $NN$ pair the use of the eikonal approximation 
is not justified,whereas at very high $Q^2$ the  energy transfer may 
be too large to consider the struck nucleon as a real one. In this case,  
after the absorbtion of the virtual photon,  
the hadron state can be viewed as a superposition of many coherently interacting states.
Obviously, this interaction can lead to  a strong cancellation
of the amplitudes and, as a result, one can expect a 
strong dilution  of the FSI effects. 
Moreover, one may expect that while the usual nuclear shadowing, 
predicted by the eikonal 
approach, becomes almost $Q^2$-independent as $Q^2$ increases, 
the suppression  of the FSI
effects due to coherent interaction of 
many intermediate hadronic states
should increase with increasing  $Q^2$,
leading   asymptotically to the vanishing FSI effects. 
The propagation of the hadronic
 debris in  the residual nucleus can be  taken into account
 by various effective theoretical models. At large values of the three-momentum transfer,
 one can modify the Glauber approach by adding another channel with an
  excited, effective hadronic state propagating trough the nucleus,
  the two-channels  generalization
 of the Glauber theory (see, e.g. \cite{fskoleliovich}).  
 Another possibility is to compute directly
 the Feynman diagram on fig.\ref{fig1} bearing in mind  
  that in the intermediate
 state the propagator of the hit nucleon in the loop integration  
 can be very  far off mass-shell so that  the 
 secondary interaction of highly virtual nucleons  (the second blob in the diagram,
fig.\ref{fig1}b)  cannot be approximated 
by the  real $NN$ scattering  amplitude. It is known that
the cross section of scattering of  virtual particles should decrease with 
 increasing  virtuality.
This is equivalent to the known fact that after a particle 
has undergone an interaction,
it should elapse a finite amount of time before it can be ready for a new one.
During this time the virtual particle propagates without any interaction. Theoretically 
this  can be     taken into account by introducing suppressing form factors at 
each end of the nucleon line in the corresponding 
Feynman diagram, which  formally 
 consists in   changing  the ejectile propagators\cite{braun}:
\be 
\frac{1}{m^2-(p+Q)^{2}}\to\frac{1}{m^2-(p+Q)^{2}}-
\frac{1}{{m^*}^2-(p+Q)^{2}}
\label{fft}
\ee 
where the subtracted term may be considered as an effective  contribution from the
excited ejectile states, which makes the rescattering contribution to  vanish at superhigh
energies (equivalent to the colour transparency effect)\footnote{A similar procedure to take into
account the virtuality in the nucleon propagator has been proposed by
F. Gross and D. Riska (Phys. Rev. {\bf C36} (1987) 1928 ) 
to restore  the gauge invariance in electro disintegration of the deuteron.}.  
 
   \begin{figure}[ht]
\raggedright
 \includegraphics[height=7.5cm]{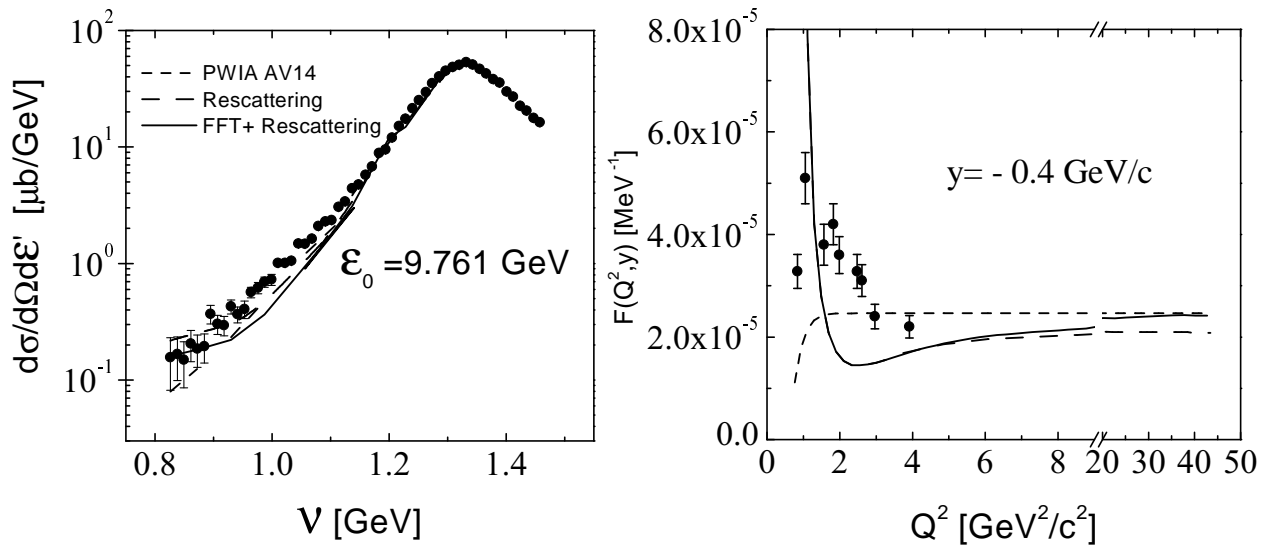}
\caption{\it The cross section (left panel) and scaling function (right panel)
for the quasi-elastic $D(e,e')X$-reactions.  The short-dashed
lines represent the results of calculations within the PWIA;
 the dashed curves correspond to the Glauber approximation;
 the solid curves are the results which include  rescattering and Finite Formation
 Time effects.
}
\label{fig4}
\end{figure}

 With the  assumption  (\ref{fft})
   the rescattering contribution  (Fig. \ref{fig1})   
  has been calculated within the assumptions
 of spinless  active nucleon, i.e. the electromagnetic
 vertex (\ref{vertex}) has been considered to  depend only on
 one elastic form factor, which can be justified
 by the fact that  in the ratios of relative contributions of rescattering
 terms to the PWIA amplitudes,  uncertainties arising from  neglecting 
 the  spin dependence of electromagnetic vertices are 
 expected to be significantly reduced and one may be able to reliable estimate the
  contribution of FFT  corrections in the total cross section. 
 Notice, that in general, the computation of  Feynman diagrams with rescattering
 effects (fig. \ref{fig1}b) produces extremely lengthy and 
  cumbersome expressions  and even in the simplest case, the
 deuteron target, a compact analytical form of results is  not feasible. 
 However, at high enough $Q^2$ and at $x_{Bj} \sim 1$ one may drastically simplify the 
 results and finally to represent the amplitude in an exactly eikonal form by formally 
 replacing the $\theta(z)$ function in (\ref{sg}) with \cite{braun}
 \be
\theta(z)\rightarrow \theta(z)\left[1-exp(\left( -i\frac{z}{l(Q^2)}\right)\right],
\label{replac}
\ee

\noindent
where $l(Q^2)=\displaystyle\frac{Q^2}{x_{Bj}mm^{* 2}}$ is the formation length.
As expected, the formation length increases with $Q^2$  and, as seen from eq. (\ref{replac}),
at $Q^2\to\infty$ it  completely cancels the contribution from the usual eikonal
rescattering term.  Note that the result (\ref{replac}) holds not only for the
deuteron target, but it is a more general result valid  
for any nuclear target, provided the multiple rescattering terms  are much smaller than
the single rescattering contribution. 
At $x_{Bj}\sim 1$ the inclusive quasi elastic reaction $D(e,e')X$
 has been studied  in Ref. \cite{braun} , whereas recently \cite{tolpa}
 the theoretical approach has been extended to the so-called cumulative region
 ($x_{Bj} > 1$) where calculations, due to the effects of GSC, which provide
   high momentum components, are much more involved.
   The results are presented 
 in Fig.\ref{fig4}.
  It can be seen that at moderate values of the  momentum  
  transfer FSI effects are visible and,
 generally, improve the description of the cross section (the PWIA, dotted curves in
 Fig.\ref{fig4}, are to be compared with rescattering contribution, dashed curves, 
 and FFT effects, the full lines). At $Q^2\to\infty$, as expected, FFT effects cancel 
 the effects of elastic rescattering and the results are close to the PWIA predictions
 (see the right panel in fig.\ref{fig4}).
 Unfortunately, present experimental data are far from the asymptotic region. 
  
\section{Conclusions}
 1. In the kinematical region of interest 
 ($y<-0.8 GeV/c$, $s$-below the pion production threshold) 
 the relativistic corrections (P-waves) 
 may be disregarded and a covariant,  relativistic 
 momentum distribution in the deuteron    may be defined which resembles
  the non relativistic  distribution. 

\noindent
 2.  FSI corrections  in inclusive 
  electro disintegration of the deuteron may be reliably estimated in a large kinematical
  region of $Q^2$ and $\nu $ by the Schroedinger, Glauber and FFT approaches

\noindent
3.  FFT effects play an important role
 in the considered reactions and, at high enough $Q^2$, 
 they lead to an almost total   cancellation of the rescattering corrections 
 predicted by  the one channel Glauber approximation. 

\acknowledgements 
This work has been done in  collaboration with 
Profs. C. Ciofi degli Atti and M.A. Braun whose contribution is
gratefully acknowledged.

\end{document}